\documentclass[aps,prd,preprint,groupedaddress,showpacs]{revtex4-1}

\usepackage{amsmath,amssymb}
\usepackage[dvips]{graphicx}
\usepackage{braket}
\newcommand{\Slash}[1]{{\ooalign{\hfil/\hfil\crcr$#1$}}}

\begin{document}

\preprint{OU-HET 774/2012}

\title{Superluminal Propagation Caused by Radiative Corrections in a\\ Uniform Electromagnetic Field}


\author{Noburo Shiba}
\email{shiba@het.phys.sci.osaka-u.ac.jp} 
\affiliation{Department of Physics, Graduate School of Science,
Osaka University, Toyonaka, Osaka 560-0043, Japan}


\date{\today}

\begin{abstract}

We consider the effect of radiative corrections 
on the maximum velocity of propagation of neutral scalar fields in a uniform electromagnetic field. 
The propagator of neutral scalar fields interacting with charged fields depends on the electromagnetic field through charged particle loops.  
The kinetic terms of the scalar fields are corrected 
and the maximum velocity of the scalar particle becomes greater or less than unity. 
We show that the maximum velocity becomes greater than unity in a simple example, a neutral scalar field coupled with two charged Dirac fields by Yukawa interaction. 
The maximum velocity depends on the frame of reference 
 and causality is not violated.
We discuss the possibility of this superluminal propagation in the Standard Model. 


\end{abstract}



\maketitle

\section{Introduction}


Classical field equations (Klein-Gordon equation, Maxwell equation, etc.) prohibit superluminal propagation. 
However, as shown in \cite{Drummond:1979pp}, Maxwell equation in curved spacetime is modified by radiative corrections 
and the photons travel at speeds greater than unity in some background gravitational field without any inconsistency. 
In this paper we consider the effect of radiative corrections on the maximum velocity of propagation of neutral fields 
in a uniform external electromagnetic field in Minkowski spacetime. 
We show that radiative corrections in a uniform electromagnetic field modify 
the kinetic terms of the field equation and the maximum velocities become greater or less than unity depending on the structure of the quantum field theory 
even in Minkowski spacetime.  
Moreover we show that the maximum velocity becomes greater than unity in a simple example, a neutral scalar field coupled with two charged Dirac fields by Yukawa interaction. 


In order to make the effect of radiative corrections as clear as possible, 
we consider propagation of a neutral real scalar field interacting with charged fields. 
The scalar field  depends on the electromagnetic field only through charged particle loops.  

First we consider generally the effect of radiative corrections 
and derive the electromagnetically modified field equation for the neutral scalar field. 
We need only consider the two points correlation function 
because we consider only the propagation of individual scalar particles.
From the translational invariance and the gauge invariance, 
the connected part of the 
propagator of the scalar field is given by, 
\begin{equation}
G(x-y,F)\equiv \bra{\Omega ,F } \mathrm{T} \phi (x) \phi (y) \ket{\Omega ,F }_c =\int \dfrac{d^4 p}{(2\pi) ^4} \dfrac{i e^{-ip(x-y)}}{p^2-\mu ^2 +\Pi (p^2)+ \tilde{\Pi} (p,F)},   \label{eq:1-1}
\end{equation}
 where $\ket{\Omega ,F }$ is the vacuum state in a uniform electromagnetic field, $F_{\mu \nu} =const$, 
 and $\bra{\Omega ,F } \mathrm{T} \phi (x) \phi (y) \ket{\Omega ,F }_c\equiv \bra{\Omega ,F } \mathrm{T} \phi (x) \phi (y) \ket{\Omega ,F }- \bra{\Omega ,F } \phi (x)  \ket{\Omega ,F } \bra{\Omega ,F } \phi (y)  \ket{\Omega ,F }$ . 
$\Pi (p^2)$ is the self energy for $F_{\mu \nu} =0$ and $\tilde{\Pi} (p,F=0)=0$. 
From the gauge invariance $G(x-y)$ depends on $F_{\mu \nu}$ rather than $A_{\mu}$.
In general, $\bra{\Omega ,F } \phi (x)  \ket{\Omega ,F } \neq 0$ when $F_{\mu \nu} \neq 0$, 
but it has no effect on the propagation and we can omit it. 
From the Lorentz covariance we obtain,  
\begin{equation}
G(\Lambda(x-y),F_{\Lambda})=G(x-y,F) ,   \label{eq:1-2}
\end{equation}
where $\Lambda$ is a Lorentz transformation, i.e. $(\Lambda(x-y))^{\mu} \equiv \Lambda^{\mu}_{\ \nu}(x-y)^{\nu}$ and 
$(F_\Lambda)^{\mu \nu}\equiv \Lambda^{\mu}_{\ \rho} \Lambda^{\nu}_{\ \sigma} F^{\rho \sigma}$. 
From (\ref{eq:1-1}) and (\ref{eq:1-2}),  $ \tilde{\Pi} (p,F)$ is a scalar function constructed by $p_{\mu}$ and $F_{\mu \nu}$. 
The general form of $\tilde{\Pi} (p,F)$ is given by
\begin{equation}
\tilde{\Pi } (p,F)=(a_1 f+a_2 h)p^2 +b_1 f+b_2 h +c_1 (F^2)^{\mu \nu} p_{\mu} p_{\nu} + O(F^4,p^4),   \label{eq:1-3}
\end{equation}
where $f\equiv - \tfrac{1}{2} F^{\mu \nu } F_{\mu \nu } =  \tfrac{1}{2} \tilde{F}^{\mu \nu } \tilde{F}_{\mu \nu } =\mathbf{E}^2 -\mathbf{B}^2 $, 
$h\equiv - \tfrac{1}{2} F^{\mu \nu } \tilde{F}_{\mu \nu } =2\mathbf{E} \cdot \mathbf{B}$, 
$\tilde{F}_{\mu \nu } \equiv \tfrac{1}{2} \epsilon _{\mu \nu \rho \sigma } F^{\rho \sigma } $, 
and $(F^n)^{\mu } _{ \  \nu }\equiv \overbrace{ F^{\mu } _{ \  \nu_1 } F^{\nu_1 } _{ \  \nu_2 } \cdots F^{\nu_{n-1} } _{ \  \nu } }^{n} $.
$\tilde{\Pi } (p,F)$ in (\ref{eq:1-3}) is the most general form because of 
the identities $(F \tilde{F} )^{\mu \nu } =\tfrac{h}{2} \eta ^{\mu \nu }$ and $ (\tilde{F}^2)^{\mu \nu } =(F^2)^{\mu \nu } -f \eta ^{\mu \nu } $. 
For simplicity, we assume that the mass ($\mu$) of the scalar field is much smaller than the mass scale ($M$) of the  charged fields ($\mu \ll M$) 
and we restrict $p_\mu$ within $| p_\mu |\ll M$. 
The higher order terms of $p_\mu$ in $\tilde{\Pi} (p,F)$ appear as $(p_\mu/M)^n$ and we neglect them. 
And we assume $|e F_{\mu \nu} |/M^2\ll 1$ and retain the lowest order terms of $F_{\mu \nu}$.
(The components $F_{\mu \nu}$ depend on the coordinate system (i.e. inertial frame), 
however, when $|h|/M^4 , |f|/M^4 \ll 1$ there are coordinate systems in which $|e F_{\mu \nu} |/M^2\ll 1$.)

We consider the shift of the pole at $p^2=\mu^2$.
From the renormalization condition $\Pi (p^2=\mu^2)=0$ and $\tfrac{d}{dp^2} \Pi (p^2)|_{p^2=\mu^2} =0$, 
so we can neglect $\Pi (p^2)$ around $p^2\approx \mu^2$.  
Then, when $p^2\approx \mu^2$ we obtain
\begin{equation}
\begin{split}
&p^2-\mu^2 +\Pi (p^2)+ \tilde{\Pi } (p,F) \approx p^2-\mu^2 + (a_1 f+a_2 h)p^2 +b_1 f+b_2 h +c_1 (F^2)^{\mu \nu} p_{\mu} p_{\nu}  \\
&\approx  (1+a_1 f+a_2 h) (p^2 +c_1 (F^2)^{\mu \nu} p_{\mu} p_{\nu} ) -(\mu^2 -b_1 f-b_2 h ) .  \label{eq:1-4}
\end{split}
\end{equation}
The dispersion relation is given by setting $(\ref{eq:1-4})$ equal to zero. 
From (\ref{eq:1-4}) we see that the modified equation of motion for the scalar field is given by
\begin{equation}
[(1+a_1 f+a_2 h) (\partial ^2 +c_1 (F^2)^{\mu \nu} \partial _{\mu} \partial _{\nu} ) +\mu'^2 ] \phi (x)=0,  \label{eq:1-5}
\end{equation}
where $\mu'^2 \equiv  \mu^2 -b_1 f-b_2 h$.
From the kinetic term in (\ref{eq:1-5}) we see that the spacetime metric for the scalar field changes as
\begin{equation}
\eta ^{\mu \nu} \rightarrow  (1+a_1 f+a_2 h) ( \eta ^{\mu \nu}+c_1 (F^2)^{\mu \nu}) \equiv (1+a_1 f+a_2 h) S^{\mu \nu}  .  \label{eq:1-6}
\end{equation}
Notice that $(1+a_1 f+a_2 h) S^{\mu \nu}$ is not universal as the spacetime metric because 
$a_1,a_2$ and $c_1$ depend on the coupling constant of the interaction between the scalar field and the charged fields.  
$(1+a_1 f+a_2 h) S^{\mu \nu}$ is considered as the spacetime metric only for the scalar field.

Next we consider the maximum velocity of the scalar particle, i.e. the group velocity ($v^{g}_i=\partial E/\partial p_i$) for large momentum ($|p_i|\gg \mu$). 
We can neglect the mass term when we consider the maximum velocity. 
The kinetic term in (\ref{eq:1-5}) determines the maximum velocity. 
We transform $F_{\mu \nu}$ into standard forms by Lorentz transformation. 
There are two cases as follows: 
(i) $\mathbf{E}$ and $\mathbf{B}$ are parallel to each other, 
(ii) $\mathbf{E}$ and $\mathbf{B}$ are mutually perpendicular and $|\mathbf{E}| =|\mathbf{B}|$. 
First we consider the case (i). 
In this case we assume $\mathbf{E}$ along the 1-axis and obtain $(F^2)^{00} =-(F^2)^{11}= \mathbf{E}^2 $, $(F^2)^{22} =(F^2)^{33}= \mathbf{B}^2 $ 
and the other components are zero.  
We substitute this $F^2$ into the kinetic terms in (\ref{eq:1-5}) and obtain 
\begin{equation}
S^{\mu \nu} \partial _{\mu} \partial _{\nu} =(1+c_1 \mathbf{E}^2) (\partial _{0}^{2} - \partial _{1}^{2} )- (1-c_1 \mathbf{B}^2) (\partial _{2}^{2} +\partial _{3}^{2} ).  \label{eq:1-7}
\end{equation} 
From (\ref{eq:1-7}) we obtain the dispersion relation $p_0=(p_1^2+\tfrac{1-c_1 \mathbf{B}^2}{1+c_1 \mathbf{E}^2} (p_2^2+p_3^2) )^{1/2} $. Here we neglect the mass. 
Then, the maximum velocities along the parallel direction to $\mathbf{E}$ ($v_{\parallel }$) and 
along the perpendicular direction to $\mathbf{E}$ ($v_{\perp  }$) are given by, 
\begin{equation}
v_{\parallel } =1,   \   
v_{\perp } = \left( \dfrac{1-c_1 \mathbf{B}^2}{1+c_1 \mathbf{E}^2} \right)^{1/2} \approx 1-\dfrac{c_1}{2} (\mathbf{E}^2 +\mathbf{B}^2) .  \label{eq:1-8}
\end{equation} 
In the case (ii) we assume $\mathbf{E}$ along the 2-axis and $\mathbf{B}$ along the 3-axis and $\mathbf{E}_2=\mathbf{B}_3=|\mathbf{E}|$. 
We obtain $(F^2)^{00}=(F^2)^{01}=(F^2)^{10}=(F^2)^{11}=\mathbf{E}^2$ and the other components are zero and
\begin{equation}
S^{\mu \nu} \partial _{\mu} \partial _{\nu} 
= (\partial _0 +\partial _1 ) ( (1+c_1 \mathbf{E}^2) \partial _{0} - (1-c_1 \mathbf{E}^2)\partial _{1})- \partial _{2}^{2} -\partial _{3}^{2} .  \label{eq:1-9}
\end{equation} 
From (\ref{eq:1-9}) we obtain the maximum velocities along the positive and negative 1-axis ($v_{1+}$ and $v_{1-}$) and along the 2,3-axis ($v_2$ and $v_3$), 
\begin{equation}
v_{1+ }=1, \ v_{1-} = \dfrac{1-c_1 \mathbf{E}^2}{1+c_1 \mathbf{E}^2} \approx 1-2c_1 \mathbf{E}^2,  \ 
v_{2} =v_{3} = \left( \dfrac{1}{1+c_1 \mathbf{E}^2} \right)^{1/2} \approx 1-\dfrac{c_1}{2} \mathbf{E}^2  .  \label{eq:1-10}
\end{equation} 
From (\ref{eq:1-8}) and (\ref{eq:1-10}), in both cases the maximum velocities become greater (less) than unity when $c_1$ is negative (positive). 
We will calculate $c_1$ in an simple example and show that $c_1$ can be negative. 

\begin{figure}
 \includegraphics[width=6cm,angle=270,clip]{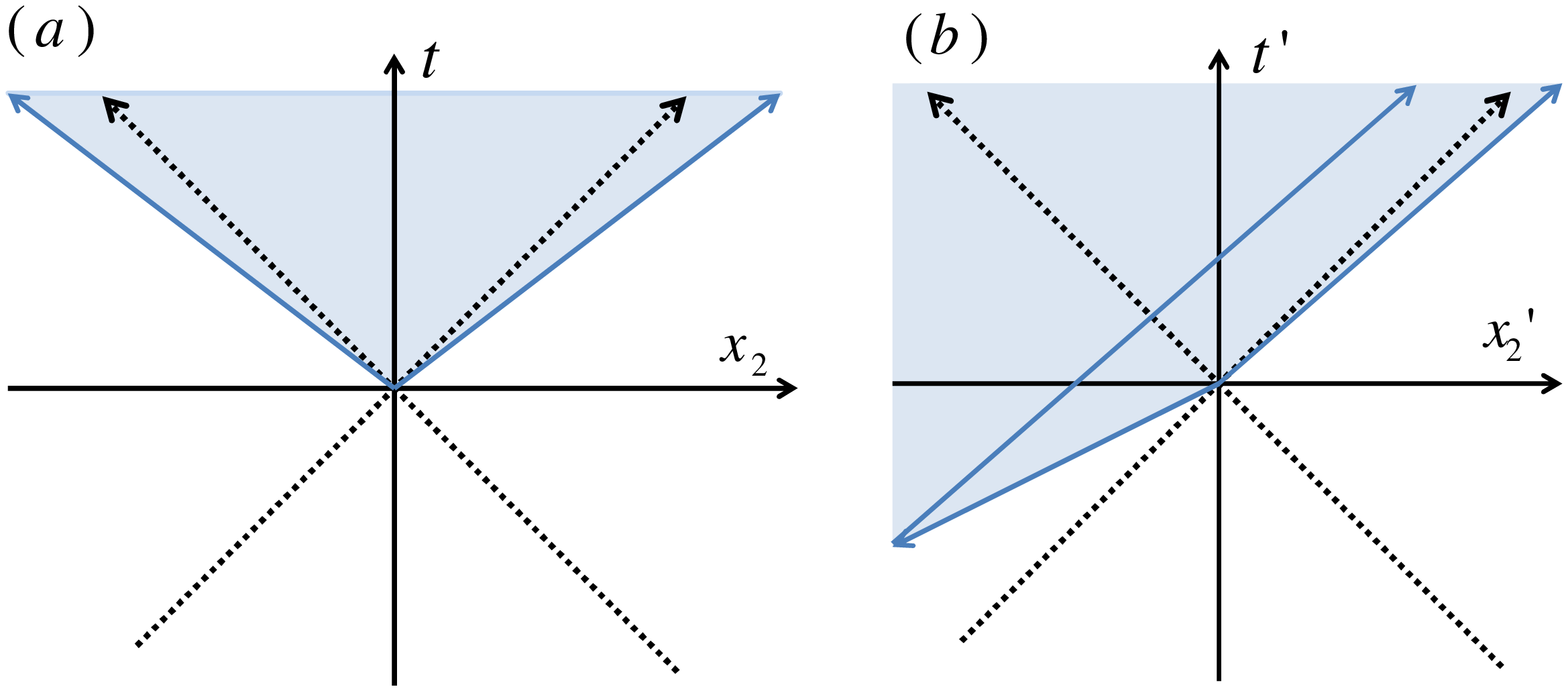}%
 \caption{ (a) The region to which an observer $A$ at $x=0$ can send a signal in the case (i).   
The dotted lines are light cones. 
(b) The same region in another inertial frame $x'$. 
A return signal back to $A$ never arrives in the past light cone of $A$ 
and causality is not violated. }
 \label{region}
 \end{figure}

Notice that causality is not violated even if the maximum velocity is greater than unity. %
The crucial point is that the maximum velocity depends on the components of $F^2$ and on the frame of reference. 
There is the assumption of relativistic invariance of the laws governing the motion 
in the usual argument about the violation of causality by faster than light transmission of information \cite{Drummond:1979pp}. 
In our case this assumption does not hold because the equation of motion depends on the components of $F^2$. 
We show the region to which a observer $A$ at $x=0$ can send a signal in the case (i) in Fig. \ref{region}(a). 
In another inertial frame $x'$ this region extends to $t'<0$ (Fig. \ref{region}(b)). 
However, from Fig. \ref{region}(b) we see that a return signal back to $A$ never arrives in the past light cone of $A$ 
and causality is not violated. 
In the same way causality is not violated also in the case (ii).

\section{A simple example}   
As a simple example, we consider a neutral real scalar field coupled with two charged Dirac fields by Yukawa interaction. 
(Our notation follows Peskin and Schroeder \cite{peskin1995introduction}.) 
We calculate explicitly $c_1$ in (\ref{eq:1-3}) in this example and show that $c_1$ can be negative.
The  Lagrangian is given by
\begin{equation}
\mathcal{L}=\dfrac{1}{2} (\partial _{\mu } \phi )^2 -\dfrac{1}{2} \mu ^2 \phi ^2 + \bar {\psi }_{1} (i \Slash{D} -m_1) \psi_{1} + \bar {\psi }_{2} (i\Slash{D} -m_2) \psi_{2} 
-g\phi (\bar {\psi }_{1} \psi_{2} +\bar {\psi }_{2} \psi_{1} ) -\dfrac{\lambda }{4!} \phi ^4  , \label{eq:2-1}
\end{equation}
where $ \psi_{1}$ and $ \psi_{2}$ have a same charge $e$ and $ D_\mu  =\partial _\mu +ie A_\mu$. 
The Lagrangian is gauge invariant and invariant under the transformation $ \phi \rightarrow - \phi, \psi_{1} \rightarrow - \psi_{1} , \psi_{2} \rightarrow \psi_{2}  $.
So the terms $ \phi , \phi^3, \phi \bar {\psi }_{1} \psi_{1} $ and $ \phi \bar {\psi }_{2} \psi_{2} $ are prohibited by the symmetry and this Lagrangian is renormalizable. 
We assume $ g^2 , e^2 , \lambda \ll 1 $ and use the perturbation theory. 
We use the dimensional regularization. 
The d-dimensional propagators of charged fermions satisfy the equation, 
\begin{equation}
(i\Slash{D} -m) S_{A}(x,x',m)=\delta ^d (x-x') , \label{eq:2-2} 
\end{equation}
where $m$ is either $m_1$ or $m_2$. 
For a uniform electromagnetic field, the exact solution of (\ref{eq:2-2}) is obtained by proper time method \cite{Schwinger:1951nm} 
(see for review \cite{itzykson2006quantum}). 
The result is 
\begin{equation}
S_{A}(x,x',m) = \Phi (x,x') (-i)  [i \Slash{\partial }_x +m] g (x-x',m) ,  \label{eq:2-3}
\end{equation}
where
\begin{equation}
\Phi (x,x') =\exp [-ie\int_{x'}^{x} d\xi ^\mu A_{\mu }(\xi ) ] \label{eq:2-4} 
\end{equation}
and
\begin{equation}
\begin{split}
g(x,m) =&\int_{0}^{\infty } ds \dfrac{i}{(4\pi i )^{d/2}} \dfrac{1}{s^{d/2}} \exp [-\dfrac{1}{2} \mathrm{tr} \ln [ (eFs)^{-1} \sinh (eFs) ] - \dfrac{i}{4} x eF \coth (eFs) x \\
&-\dfrac{i}{2} e F^{\mu \nu } \sigma _{\mu \nu } s -i(m^2-i\epsilon )s] ,  \label{eq:2-5}
\end{split}
\end{equation}
where $\sigma _{\mu \nu } =\tfrac{i}{2} [\gamma _{\mu } , \gamma _{\nu } ]$. 
The path of $\xi$ integral in the phase factor $\Phi (x,x')$ is a straight line from $x'$ to $x$. 
The role of $\Phi (x,x')$ is to make $S_A (x,x')$ gauge covariant, i.e. when 
$A_\mu (x) \rightarrow A_\mu (x)-\tfrac{1}{e} \partial _\mu \alpha (x) $,  $\Phi (x,x') \rightarrow e^{i(\alpha (x)-\alpha (x'))} \Phi (x,x')$.
In (\ref{eq:2-5}) we use following matrix notations; $ \mathrm{tr} F^n =(F^n)^{\mu}_{\ \mu} $  
and $x F^n x =x_\mu (F^n)^{\mu \nu } x_\nu $. 

For loop calculations, it is convenient to use the momentum representation of $S_{A}$ \cite{Tsai:1974ap}, 
\begin{equation}
S_{A}(x,x',m) = \Phi (x,x') (-i) \int \dfrac{d^d p}{(2\pi) ^d} e^{-ip(x-x')} [\Slash{p} +m]  \tilde{g} (p,m) ,  \label{eq:2-6} 
\end{equation}
where
\begin{equation}
\begin{split}
&(-i)\tilde{g} (p,m) =(-i)\int_{0}^{\infty } ds \exp [-\dfrac{1}{2} \mathrm{tr} \ln \cosh (eFs) + i p (eF)^{-1} \tanh (eFs) p -\dfrac{i}{2} e F^{\mu \nu } \sigma _{\mu \nu } s -i(m^2-i\epsilon )s] \\
&=\dfrac{1}{p^2-m^2 +i\epsilon } + \dfrac{1}{2} \dfrac{e F^{\mu \nu } \sigma _{\mu \nu }}{ (p^2-m^2 +i\epsilon )^2 } -\dfrac{e^2}{4} \dfrac{F_{\mu \nu } 
F_{\lambda \kappa } \gamma ^{[\mu } \gamma ^{\nu } \gamma ^{\lambda } \gamma ^{\kappa ]} }{ (p^2-m^2 +i\epsilon) ^3 } - \dfrac{2e^2 pF^2 p}{ (p^2-m^2 +i\epsilon) ^4 } +O(F^3) . 
\label{eq:2-7} 
\end{split}
\end{equation}
Here 
$\gamma ^{[\mu } \gamma ^{\nu } \gamma ^{\lambda } \gamma ^{\kappa ]} $ is the completely antisymmetrized product (for $d=4$,  
$\gamma ^{[\mu } \gamma ^{\nu } \gamma ^{\lambda } \gamma ^{\kappa ]} =-i \epsilon ^{\mu \nu \lambda \kappa } \gamma _5$).
To obtain the second line in (\ref{eq:2-7}) we have used the identity, 
$\tfrac{1}{2} \{ \sigma _{\mu \nu } ,\sigma_{\lambda \kappa } \} =\eta _{\mu \lambda } \eta _{\nu \kappa } -\eta _{\mu \kappa  } \eta _{\nu \lambda  } -\gamma _{[\mu } \gamma _{\nu } 
\gamma _{\lambda } \gamma _{\kappa ]}$. 
Because of the relation $\Phi (x,x') \Phi (x',x) =1$, the phase factor $\Phi (x,x') $ is always canceled in charged fermion loops. 
So the only difference between the ordinary Feynman rules and those in an uniform external electromagnetic field is 
the propagator. 
We need to use the propagator $(\Slash{p} +m) \tilde{g} (p,m)$ instead of $i(\Slash{p} +m)/(p^2-m^2)$.

\begin{figure}
 \includegraphics[width=4cm,angle=270,clip]{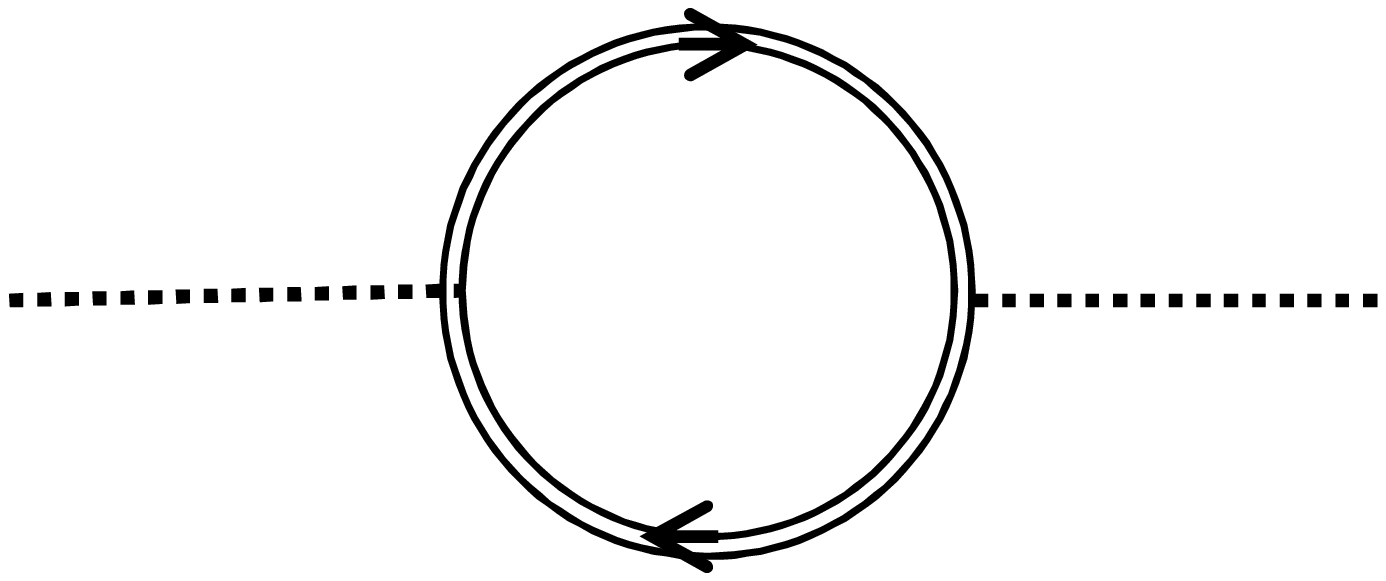}%
 \caption{ The lowest order self energy diagram which depends on $F_{\mu \nu}$. 
 The double line stands for the propagator $(\not p +m) \tilde{g} (p,m)$ in (\ref{eq:2-7}). }
 \label{diagram}
 \end{figure}

\begin{figure}
 \includegraphics[width=8cm,clip]{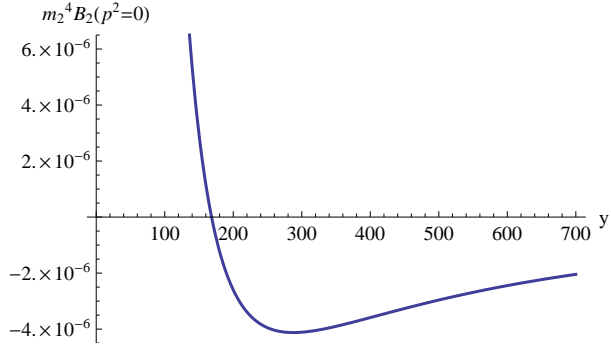}%
 \caption{ $ m_{2}^{4} B_2 (p^2=0) $ in (\ref{eq:2-14}) as a function of $y\equiv m_1^2/m_2^2$. 
$c_1=\tfrac{g^2 e^2}{(4\pi)^2} \tfrac{4}{3}  B_2(p^2=0) $ can be both positive and negative when $y$ varies. }
 \label{b2}
 \end{figure}

Fig. \ref{diagram} is the lowest order self energy diagram which depends on $F_{\mu \nu}$  
\begin{equation}
i\Pi (p,F) =-(-ig)^2 \int \dfrac{d^d k}{(2\pi)^d } \mathrm{Tr} [ (\Slash{k} +m_2 ) \tilde{g} (k,m_2) (\Slash{p} +\Slash{k} +m_1 ) \tilde{g} (p+k,m_1) ] ,  \label{eq:2-8} 
\end{equation}
where $ \mathrm{Tr} $ means the trace of $\gamma $ matrices. 
We substitute the expansion of $\tilde{g} (p,m)$ in (\ref{eq:2-7}) into (\ref{eq:2-8}) and calculate $\Pi (p,F)$ to $O(F^2)$.  
As we argued 
just above eq. (\ref{eq:1-4}), 
we consider only the terms which depend on $F_{\mu \nu}$ 
because the term which is independent of $F_{\mu \nu}$ does not contribute to the shift of the pole of the propagator at $p^2=\mu^2$. 
We can calculate $\Pi (p,F)$ by the usual techniques for loop calculations \cite{peskin1995introduction}.
The term linear in $F_{\mu \nu}$ vanishes because Lorentz invariants $F^{\mu}_{\ \mu}$ and $F^{\mu \nu} p_{\mu} p_{\nu}$ are zero. 
Because the term quadratic in $F_{\mu \nu}$ is finite at $d=4$, 
we will henceforth assume that the regularization has been removed and $d=4$. 
To calculate the term quadratic in $F_{\mu \nu}$, we use the identity 
\begin{equation}
\mathrm{Tr} [\gamma ^\mu F \cdot \sigma  \gamma ^\nu  F \cdot  \sigma  ] =2^5 (F^2)^{\mu \nu } -2^4 f \eta ^{\mu \nu } ,  \label{eq:2-9} 
\end{equation}
where $F \cdot  \sigma \equiv F^{\alpha \beta } \sigma _{\alpha \beta }$ and $f = - \tfrac{1}{2} F^{\mu \nu } F_{\mu \nu } $. 
After some calculation, we obtain the term quadratic in $F_{\mu \nu}$ 
\begin{equation}
\Pi (p,F) |_{F^2} =\dfrac{ g^2 e^2}{(4\pi)^2} \dfrac{4}{3} [f B_1(p^2) +(F^2)^{\mu \nu} p_{\mu} p_{\nu}  B_2(p^2) ] ,  \label{eq:2-10} 
\end{equation}
where
\begin{equation}
B_1=\int_{0}^{1} dx [\dfrac{1}{\Delta }(9x(1-x) -2) +\dfrac{1}{\Delta ^2} (-m_1^2x-m_2^2 (1-x) +m_1 m_2)] ,  \label{eq:2-11} 
\end{equation}
\begin{equation}
B_2=\int_{0}^{1} dx [\dfrac{1}{\Delta ^2 } (1-x)x(10 x(1-x) -1) -\dfrac{2}{\Delta ^3} (1-x)^2 x^2  (-m_1^2x-m_2^2 (1-x) +m_1 m_2)] ,  \label{eq:2-12} 
\end{equation}
and $\Delta \equiv p^2 x(x-1) +m_{1}^{2} x+m_{2}^{2} (1-x)$.
By comparing (\ref{eq:1-3}) and (\ref{eq:2-10}), we obtain the coefficients in (\ref{eq:1-3}); 
$a_2=b_2=0$, $a_1=\tfrac{g^2 e^2}{(4\pi)^2} \tfrac{4}{3} \tfrac{d B_1}{d p^2}|_{p^2=0}$,  
$b_1=\tfrac{g^2 e^2}{(4\pi)^2} \tfrac{4}{3}  B_1(p^2=0)$, and 
\begin{equation}
c_1=\dfrac{g^2 e^2}{(4\pi)^2} \dfrac{4}{3}  B_2(p^2=0) .  \label{eq:2-13} 
\end{equation}
We can calculate the $x$ integral in $B_2(p^2=0)$ and obtain 
\begin{equation} 
\begin{split}
&m_{2}^{4} B_{2} (p^2=0 ) = \dfrac{4}{(y-1)^5} [(y-1) (y^2+10y+1)-6y(y+1)\ln y] \\ 
&+\dfrac{1}{(y-1)^3} [2(y-1) -(y+1) \ln y] 
 + \dfrac{ 2\sqrt{y} }{(y-1)^5} [3(y^2-1) -(y^2+4y+1) \ln y] ,  \label{eq:2-14} 
\end{split}
\end{equation}
where $y\equiv m_1^2/m_2^2$. 
In particular, we notice that $ m_{2}^{4} B_2 (p^2=0 ,y=1) =1/6$ and $m_{2}^{4} B_2 (p^2=0 ,y\rightarrow \infty ) \approx -(\ln y)/y^2$ 
(we assume $m_1\geq m_2$). 
So $c_1$ in (\ref{eq:2-13}) as a function of $y$ is negative for large $y$. 
We show the graph of $ m_{2}^{4} B_2 (p^2=0 ,y) $ as a function of $y$ in Fig. \ref{b2}, 
and we see that $c_1$ can be both positive and negative when $y$ varies. 
From (\ref{eq:1-8}) and (\ref{eq:1-10}), the maximum velocities of the scalar particle become greater than unity when $c_1$ is negative,  
so the scalar particle in this example can propagate at speed  greater than unity in an uniform electromagnetic field.

\section{Conclusion and discussion} 
In this paper we studied the maximum velocity of a neutral real scalar field interacting with charged fields in an uniform electromagnetic field. 
We showed that the kinetic terms of the scalar fields are modified by the radiative corrections  
and the maximum velocity of the scalar particle becomes greater or less than unity depending on the structure of the quantum field theory.  
From (\ref{eq:1-8}) and (\ref{eq:1-10}), the maximum velocity become greater (less) than unity when $c_1$ in (\ref{eq:1-3}) is negative (positive).
The physical interpretation of this effect is that the vacuum is considered as the medium filled with imaginary charged particles 
and the external electromagnetic field changes the property of the vacuum. 
The maximum velocity depends on the frame of reference (i.e. the components of $F_{\mu \nu}$) 
 and causality is not violated even if the maximum velocity is greater than unity. 
From (\ref{eq:1-6}) this change of the kinetic term can be considered as the change of the spacetime metric for the scalar field. 
So the change of the kinetic term also modifies probably the lifetime of the particle in an electromagnetic field.

We considered the neutral real scalar field in an uniform electromagnetic field to make the effect of radiative corrections clear. 
We can generalize the analysis in various ways. 
For example, the maximum velocity of charged fields and of higher spin fields becomes probably greater or less than unity 
depending on the structure of the quantum field theory
in an uniform electromagnetic field.

Finally we discuss the possibility of this superluminal propagation in the Standard Model. 
The Yukawa coupling of mesons is greater than one and the perturbative calculation is not reliable. 
So we cannot know the sign of $c_1$,  
however, by dimensional analysis we can estimate the magnitude of $c_1$; $|c_1|\approx e^2 m_p^{-4} $, 
where $m_p$ is the proton mass. 
From (\ref{eq:1-8}) the shift of the maximum velocity from unity in an uniform electromagnetic field that $\mathbf{E}$ and $\mathbf{B}$ are parallel to each other is 
$|v_{\perp }-1|\approx e^{2} m_p^{-4} (\mathbf{B}^2 + \mathbf{E}^2) $. 
As an example, we estimate this value for the electric field $\mathbf{E}_{n}$ near a nucleus ($e |\mathbf{E}_{n}| \approx e^2 m_p^2 $): 
$|v_{\perp }-1|\approx e^{4} \approx 10^{-4} $. 
We can possibly observe the change of the maximum velocity in the propagation of light neutral mesons in matter.

A simple case in which the perturbative calculation is reliable is the case of neutrinos \cite{Shiban}. 
The propagator of neutrinos depends on the electromagnetic field through $W$ boson charged lepton loops. 
The mechanism studied in this paper can possibly explain the experiment by MINOS collaboration \cite{Adamson:2007zzb}  
which reported that the speed of neutrinos is greater than unity. 

We have not had an intuitive explanation for the negative sign of $c_1$. 
For example, we can see easily that the sign of $c_1$ for a neutral real scalar field interacting with two charged scalar fields which have different masses is always positive.  
So the difference of masses of charged fields is not a sufficient condition. 
We need extensive 
studies to know whether there is the superluminal propagation in the Standard Model or not. 

\begin{acknowledgments}
I am grateful to Takahiro Kubota  for a careful reading of 
this manuscript and useful comments and discussions. 
I also thank Yutaka Hosotani for useful discussions.
This work was supported
by a Grant-in-Aid from JSPS (No. 22-1930).
\end{acknowledgments}

\end{document}